# Magnetic order in the quasi-one-dimensional Ising system RbCoCl$_3$

N P Hänni,[1,2] D Sheptyakov,[3] M Mena,[3,4] E Hirtenlechner,[3,5] L Keller,[3] U Stuhr,[3] L-P Regnault,[5] M Medarde,[6] A Cervellino,[7] Ch Rüegg,[8,9,10,11] B Normand[8,11] and K W Krämer[1]

[1] Department for Chemistry and Biochemistry, University of Bern, Freiestrasse 3, CH-3012 Bern, Switzerland
[2] Space Research and Planetary Sciences, Physics Institute, University of Bern, Sidlerstrasse 5, CH-3012 Bern, Switzerland
[3] Laboratory for Neutron Scattering and Imaging, Paul Scherrer Institute, CH-5232 Villigen-PSI, Switzerland
[4] London Centre for Nanotechnology and Department of Physics and Astronomy, University College London, Gower Street, London WC1E 6BT, United Kingdom
[5] Institut Laue-Langevin, 71 avenue des Martyrs, CS 20156, 38042 Grenoble Cedex, France
[6] Laboratory for Multiscale Materials Experiments, Paul Scherrer Institute, CH-5232 Villigen-PSI, Switzerland
[7] Swiss Light Source, Paul Scherrer Institute, CH-5232 Villigen-PSI, Switzerland
[8] Paul Scherrer Institute, CH-5232 Villigen-PSI, Switzerland
[9] Department of Quantum Matter Physics, University of Geneva, CH-1211 Geneva, Switzerland
[10] Institute for Quantum Electronics, ETH Zürich, CH-8093 Zürich, Switzerland
[11] Institute of Physics, Ecole Polytechnique Fédérale de Lausanne, CH-1015 Lausanne, Switzerland

## Abstract

The dynamical properties of free and bound domain-wall excitations in Ising-chain materials have recently become the focus of intense research interest. New materials and spectrometers have made it possible to control the environment of coupled Ising chains by both effective internal and applied external fields, which can be both longitudinal and transverse, and thus to demonstrate how the resulting magnetic phase transitions and the nature of the associated excited states obey fundamental symmetry properties. In RbCoCl$_3$, the weakly coupled Ising chains form a triangular lattice whose frustrated geometry and magnetic ordering transitions at low temperature open new possibilities for the Ising-chain environment. We have investigated the structure and magnetism in RbCoCl$_3$ by high-resolution x-ray diffraction and neutron scattering measurements on powder and single crystal samples between 1.5 K and 300 K. Upon cooling, the Co$^{2+}$ spins develop one-dimensional antiferromagnetic correlations along the chain axis ($c$-axis) below 90 K. Below the first Néel temperature, $T_{N1}$ = 28 K, a partial 3D magnetic order sets in, with propagation vector $\mathbf{k_1}$ = (1/3,1/3,1), the moments aligned along the $c$-axis and every third chain uncorrelated from its neighbours. Only below a second magnetic phase transition at $T_{N2}$ = 13 K does the system achieve a fully ordered state, with two additional propagation vectors: $\mathbf{k_2}$ = (0,0,1) establishes a "honeycomb" $c$-axis order, in which 1/3 of the chains are subject to a strong effective mean field due to their neighbours whereas 2/3 experience no net field, while $\mathbf{k_3}$ = (1/2,0,1) governs a small, staggered in-plane ordered moment. We conclude that RbCoCl$_3$ is an excellent material to study the physics of Ising chains in a wide variety of temperature-controlled environments and our results have an important role in interpreting measurements of the spin dynamics.

# 1. Introduction

The Ising model describes quantum spins coupled by only one of their components, and in one dimension it can be solved exactly [1], including in the presence of a longitudinal magnetic field, to obtain all of its thermodynamic and dynamic properties. Its spin excitations, domain walls separating ordered regions, are formed in pairs by a single spin-flip and thus present a simple paradigm for half-integer magnetic excitations, which in more complex quantum models develop into solitons and spinons. In a transverse field the Ising model is also exactly soluble [2], providing further foundation stones for the physics of non-commuting quantum operators, duality and also topology.

Although compounds realising the one-dimensional (1D) Ising model have long been known, and the continuum scattering characteristic of fractional excitations was observed by inelastic neutron scattering in the early 1980s [3,4,5], recent experiments on a new generation of Ising-chain materials have revealed surprisingly rich new physics, and in particular new dynamical phenomena. Depending on the precise symmetry of the intrachain interactions (gauged by the parameter $\varepsilon$, where $\varepsilon = 0$ is pure Ising and $\varepsilon = 1$ is Heisenberg) and on the geometry of the weak interchain interactions, one may find complex patterns of both continua and bound states with characteristic polarisation properties. $CoNbO_6$ has chains of strongly Ising character ($\varepsilon = 0.1$) with ferromagnetic (FM) interactions and its spectrum consists of a Zeeman ladder of (anti-)bound states that shows the symmetry of the $E_8$ Lie group at the transverse-field-tuned quantum critical point [6]. The materials $ACo_2V_2O_8$ (A = Sr, Ba) realise antiferromagnetic (AFM) chains with weaker Ising anisotropy ($\varepsilon = 0.5$), whose spiralling geometry allows the application of both uniform and staggered magnetic fields, both sufficiently large to drive the quantum phase transition [7,8], and whose dynamical properties exhibit both longitudinal and transverse excitations [9]. In $CoNbO_6$, subsequent experiments have revealed further complex excited states by inelastic neutron scattering (INS) [10,11] and terahertz spectroscopy [12]. In $ACo_2V_2O_8$, further investigation has revealed spinon confinement and bound-state formation [13,14], emergent fermions in a transverse field [15], Bethe strings [16] and topological excitations [17].

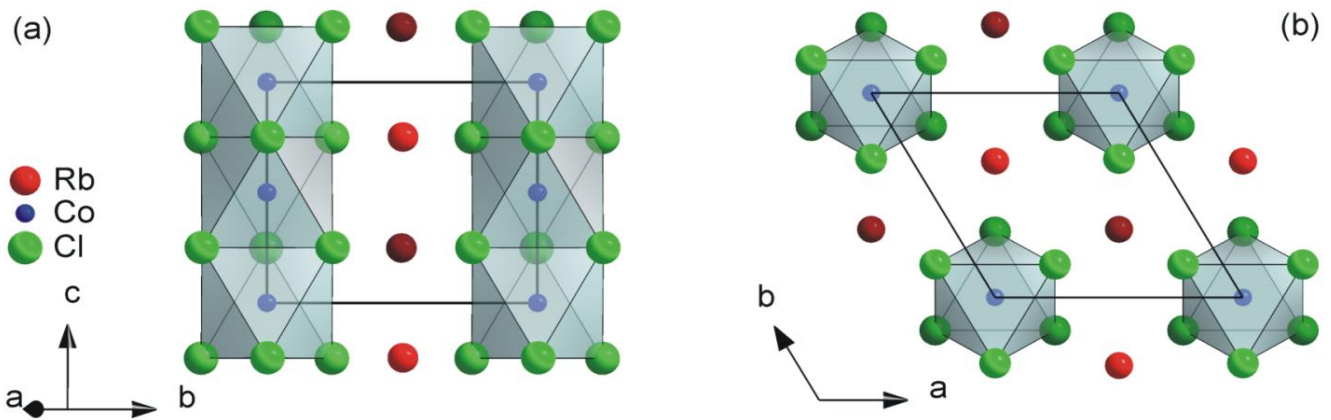

**Figure 1.** Crystal structure of the hexagonal perovskite $RbCoCl_3$. (a) Face-sharing $[CoCl_{6/2}]^-$ octahedra form chains along the $c$-axis. (b) View onto the hexagonal plane.

From this perspective, it is time to revisit the hexagonal AMX$_3$ perovskite materials. AMX$_3$ systems with M = Co$^{2+}$ as the magnetic ion, X = (Cl$^-$, Br$^-$) and A a monovalent cation form good Ising chains, in which the early observations of the domain-wall continuum were made [3,4,5]. With a renewed focus on intrachain Heisenberg interactions and confining or effective-field interchain interactions, hexagonal AMX$_3$ compounds offer a wide range of possibilities for elemental substitution and a triangular geometry of chains whose frustration presents a number of low-temperature magnetic structures. As an essential accompaniment to spectroscopic studies of bound and deconfined domain-wall excitations in the AFM Ising-chain materials of the AMX$_3$ class, here we perform a detailed investigation of these low-temperature magnetic structures in one of its key representatives, RbCoCl$_3$.

**Table 1.** Magnetic structures of selected AMX$_3$ compounds. The propagation vector, **k**, and the moment directions are given with respect to the CsNiCl$_3$-type unit cell (P6$_3$/mmc). $V_{mag}/V$ is the increase of the volume of the magnetic unit cell and $T_N$ the Néel temperature.

| compound | **k**-vector | $V_{mag}/V$ | moment direction | magnetic structure $ab$-plane | magnetic structure $c$-axis | $T_N$ (K) | μ (μ$_B$) | Ref. |
|---|---|---|---|---|---|---|---|---|
| CsVCl$_3$ | (1/3,1/3,1) | 3 | xy | 120° AFM | AFM | 14 | 1.97 | [23] |
| RbVBr$_3$ [a] | (1/3,1/3,1) | 3 | xy | FIM | AFM | 28 | 1.97 | [23-26] |
|  |  |  | xy | 120° AFM | AFM | 21 |  |  |
| CsCrCl$_3$ [b] | (1/2,0,1) | 2 | xy | AFM | AFM | 16 | 3.2 | [28] |
| CsMnBr$_3$ | (1/3,1/3,1) | 3 | xy | 120° AFM | AFM | 8.3 | 3.3 | [32,33] |
| CsMnI$_3$ | (1/3,1/3,1) | 3 | z | FIM | AFM | 11.4 | 3.60 | [34,35] |
|  |  |  | xz | 51° FIM | AFM | 8.2 |  |  |
| CsFeCl$_3$ | (1/3,1/3,0) | 3 | xy | 120° AFM | FM | 2.6 | 1.3 | [40,41] |
| TlCoCl$_3$ [a] | [c] | 8 | z | AFM | AFM | 29.5 | 2.7 | [42] |
| CsCoCl$_3$ | (1/3,1/3,1) | 3 | z | FIM | AFM | 20.8 | 3.1 | [43,44] |
|  |  |  |  |  |  | 13.5 |  |  |
| RbCoBr$_3$ [b] | (1/3,1/3,1) | 3 | z | FIM | AFM | 37.2 |  | [45] |
|  |  |  |  |  |  | 30.5 |  |  |
| CsCoBr$_3$ | [d] |  6 | z | AFM | AFM | 28 | 3.0 | [46] |
|  |  |  | xz |  | AFM | 12 |  |  |
| CsNiCl$_3$ | (1/3,1/3,1) | 3 | z | FIM | AFM | 4.8 | 1.0 | [36,37] |
|  |  |  | xz | 120° AFM | AFM | 4.4 |  |  |
| RbNiCl$_3$ | (1/3,1/3,1) | 3 | xz | 57.5° FIM | AFM | 11.1 | 1.3 | [38,39] |
| CsCuCl$_3$ [b] | (1/3,1/3,0.014) | (3) | xy | 120° AFM | 5.1° spiral | 10.7 | 0.58 | [29] |

a) KNiCl$_3$-type lattice distortions.
b) Jahn-Teller distortion. Values are given with respect to the hexagonal parent cell.
c) $4a_0$ × $\sqrt{3}a_0$ × $c_0$ magnetic cell.
d) $\sqrt{3}a_0$ × $3a_0$ × $c_0$ magnetic cell.

More generally, AMX$_3$ perovskites have been the focus of numerous investigations over the past 50 years to characterise their structures, structural phase transitions, chemical, magnetic and spectroscopic properties. Within this broad class of materials, hexagonal perovskites with the 2H structure (space group

P6$_3$/mmc) take a prominent position due to their quasi-1D structure, in which chains of face-sharing octahedra [MX$_{6/2}$] are well separated from neighbouring chains by the A cations, as illustrated in figure 1. In comparison to corner- or edge-sharing polyhedra, face-sharing octahedra provide the shortest possible M-M distance and therefore an optimal basis for cooperative, low-dimensional physical properties, including the AFM Ising-chain character of RbCoCl$_3$.

Many magnetic structures have been determined for the series of 2H AMX$_3$ compounds in which A$^+$ is Rb$^+$ or Cs$^+$, M$^{2+}$ is a 3d transition-metal ion and X$^-$ is a halide ion, with selected examples summarised in Table 1. Their magnetic order is a sometimes complex compromise governed by the crystal structure, the single-ion electronic properties and the interactions between magnetic ions. In space group P6$_3$/mmc, the M$^{2+}$ cation on site (2a) has point symmetry -3m (D$_{3d}$ in Schönflies notation), a trigonally distorted octahedral coordination that splits the $t_{2g}$ orbitals into low-lying $A_{1g}$ and $E_g$ states, whereas the $e_g$ orbitals remain in another $E_g$ state. For d$^2$, d$^4$, d$^7$ and d$^9$ transition-metal ions with trigonal symmetry, the E$_g$ states may be further split by the Jahn-Teller (JT) effect, whereas d$^3$, d$^5$ and d$^8$ ions do not show JT splitting and only a small, trigonal single-ion anisotropy. The magnetic interactions in 2H AMX$_3$ compounds are strongly spatially anisotropic, with the intrachain coupling constants, $J$, exceeding the interchain couplings, $J'$, by at least an order of magnitude.

The magnetic unit cell of most 2H AMX$_3$ compounds is tripled in the hexagonal plane. When the interactions are isotropic in spin space (Heisenberg-type), one magnetic phase transition is observed and the ordered moments are located in the hexagonal plane with a 120° AFM structure [18]. By contrast, for systems with Ising-type anisotropy one finds two successive phase transitions as the temperature is lowered. First the moments adopt a ferrimagnetic (FIM) order with collinear up-down-up alignment along the $c$-axis (the chain axis), and then the components lying in the hexagonal plane order at a second transition to align the moments in the $ac$-plane with a near-120° AFM spin structure [19].

For further context we review the magnetic order of related AMCl$_3$ compounds, where A is mostly Cs. The compounds CsSc$_{1-x}$Cl$_3$ [20,21] (d$^1$) and CsTiCl$_3$ [22] (d$^2$) are known, but no magnetic investigations have been published. The V$^{2+}$ compounds (d$^3$) have a small $xy$ (spatial) anisotropy without JT distortion. As expected for a Heisenberg coupling, CsVCl$_3$ shows a single magnetic phase transition from short-ranged 1D correlations to a 3D ordered state with a tripled magnetic unit cell [23] and the moments located on 120° AFM sublattices in the $ab$-plane, AFM coupled along the $c$-axis. However, for smaller r(A)/r(X) ratios, where r(A) and r(X) are the respective ionic radii of A and X, KNiF$_3$-type lattice distortions occur; the unequal [MX$_{6/2}$] chains have different magnetic interactions and in the case of RbVBr$_3$ this leads to two magnetic phase transitions [24,25], with the intermediate phase showing a collinear FIM order [26]. Both CsCrCl$_3$ (d$^4$) and CsCuCl$_3$ (d$^9$) have Heisenberg coupling and show structural distortions due to the JT effect, thus failing to conform to the general picture. For CsCrCl$_3$, the volume of the hexagonal cell is doubled towards a C-centred monoclinic cell with space group C2/m [27], which is also the magnetic unit cell [28]. CsCuCl$_3$ has an incommensurate magnetic structure with 120° AFM order in the $ab$-plane and a 5.1° spiral along the $c$-axis [29]. For Mn$^{2+}$ compounds (d$^5$) the situation is rather complex, despite the absence of a JT splitting, the small single-ion anisotropy and the Heisenberg coupling. The chlorides adopt two different types of crystal structure, as in the case of CsMnCl$_3$ 9R [30] and RbMnCl$_3$ 6H [31], which prohibits a comparison with the 2H compounds. CsMnBr$_3$ behaves in the same way as the generic CsVCl$_3$ example [32,33], but a decreasing r(A)/r(X) ratio causes the anisotropy to increase and CsMnI$_3$ shows two successive magnetic phase transitions with moments in the *ac*-plane, as for an Ising-like Heisenberg AFM [19]. The second phase has a near-120° AFM spin arrangement in the *ac*-plane, but the angles to the *c*-axis deviate from 60° for two of the spins and a small FIM moment remains [34,35]. The $Ni^{2+}$ compounds ($d^8$) behave very similarly, $CsNiCl_3$ showing two transitions that are very close in temperature [36,37], and merge into one for $RbNiCl_3$, which has a tiny Ising anisotropy [38,39]. $CsFeCl_3$ ($d^6$) has a singlet ground-state and shows magnetic order only in an external field ($H \parallel c > 4$ T); uniquely, this compound has a FM coupling along the *c*-axis [40,41]. Finally, among the $Co^{2+}$ compounds ($d^7$) $TlCoCl_3$ shows $KNiCl_3$-type lattice distortions [42], whereas $CsCoCl_3$ [43,44], $RbCoBr_3$ [45] and $CsCoBr_3$ [46] all maintain the higher structural symmetry and exhibit two successive magnetic phase transitions with the moments remaining aligned along the *c*-axis.

We chose to investigate $RbCoCl_3$ with the aim of finding a generic Ising spin-chain material having the lowest magnetic interactions in the $AMX_3$ class and no additional structural distortions. To date the structure of the magnetically ordered phases of $RbCoCl_3$ has not been determined, it being known from Raman [47,48], infrared [49] and Mössbauer [50] spectroscopy only that three distinct regimes of magnetic behaviour appear, in analogy to $CsCoCl_3$. Here we fill this gap by reporting the full details of the crystal structure, the magnetic phases and their spin alignments in $RbCoCl_3$, thereby providing a sound basis for the accurate interpretation of spectroscopic measurements [51].

## 2. Material and Methods

$RbCoCl_3$ was synthesised from RbCl and $CoCl_2$. The RbCl (Merck, suprapur) was dried at 200°C under vacuum and the $CoCl_2$ was obtained by dehydration of $CoCl_2 \cdot 6 H_2O$ (Strem, 5N). The hydrate was heated by 10 °C/h to 90°C in an argon gas flow and then to 200°C under vacuum. The dry $CoCl_2$ was sublimed under vacuum at 560°C to remove traces of oxide impurities. Stoichiometric amounts of the starting materials were then sealed in silica ampoules under vacuum and crystals were grown by the Bridgman technique using a vertically moving furnace with a temperature gradient. The material was molten at 505°C and solidification took place from the lower tip of the ampoule as the furnace was moved slowly upwards (0.01 mm/min). All starting materials and products were kept under dry and oxygen-free conditions ($H_2O$ and $O_2$ levels below 0.1 ppm) in glove boxes or sealed ampoules and sample containers. As a diamagnetic reference material for the specific-heat measurements, $CsMgCl_3$ was prepared from CsCl (Merck, suprapur) and $MgCl_2$ (Cerac, 3N). The CsCl was dried at 200°C under vacuum and the $MgCl_2$ was sublimed at 750°C under vacuum. For crystal growth by the Bridgman technique, the stoichiometric mixture of the starting materials was molten at 625°C.

The phase purity of the ternary chlorides was verified by x-ray powder diffraction on a STOE StadiP diffractometer in Bragg-Brentano geometry using Cu $K_{\alpha 1}$ radiation monochromated by α-quartz. The melting point of $RbCoCl_3$ was determined by differential scanning calorimetry on a Mettler DSC 823e in a cold-welded gold crucible and our result, 486°C, agrees with the literature value of 485°C [52].

Specific-heat measurements were performed on a Quantum Design Physical Property Measurement System (PPMS). Small, plate-like single crystals of $RbCoCl_3$ (of approximate dimensions 1.8 x 1.8 x 0.3 $mm^3$; 9.7 mg) and $CsMgCl_3$ (1.4 x 1.6 x 0.2 $mm^3$; 8.3 mg) were mounted on a sapphire plate using a small amount of Apiezon N grease. Measurements were made between 2 K and 300 K under identical conditions both for the samples and for the empty sample holder with the grease.

The magnetic properties were measured on a Quantum Design MPMS-5XL SQUID magnetometer. A finely ground powder (20.1 mg) and two single crystals of RbCoCl$_3$ (2.0 x 4.0 x 0.6 mm$^3$; 74.3 mg and 2.0 x 2.0 x 0.4 mm$^3$; 16.9 mg) were placed in gelatine capsules using a small amount of Apiezon N for fixation, the first crystal with the *c*-axis orientated parallel and the second with the *c*-axis orientated perpendicular to the magnetic field. The magnetic susceptibility, $\chi(T)$, was measured in the RSO mode while cooling from 300 K to 1.9 K in a field of 0.05 T and the magnetisation, $M(H)$, at 1.9 K in fields up to 5 T. The susceptibility data were corrected for the diamagnetic contributions of RbCoCl$_3$, calculated from Pascal's constants [53] to be -1.1x10$^{-4}$ cm$^3$mol$^{-1}$, and of the empty sample holder.

Crystallites of RbCoCl$_3$ were ground finely to obtain powder samples for our neutron diffraction experiments. A vanadium sample container of 8 mm diameter and 50 mm length was filled with powder under a helium atmosphere. Neutron diffraction patterns were measured on the powder diffractometers HRPT [54] and DMC [55] at the Swiss Spallation Neutron Source (SINQ) at the Paul Scherrer Institute (PSI) in Villigen, Switzerland. Data were collected between 1.5 K and 300 K using ILL-type helium cryostats and neutron wavelengths ($\lambda$) of 1.154 Å, 1.494 Å and 2.450 Å on HRPT and 2.455 Å on DMC. Diffraction data from HRPT were used for structural refinements and from DMC for magnetic refinements. Powder diffraction data were also collected with a Mythen II detector on the powder station (MS-powder) of the Material Sciences beamline [56] at the Swiss Light Source (SLS, PSI). Samples in 0.3 mm glass capillaries were measured using synchrotron radiation of wavelength 0.62041 Å, with temperatures between 4 K and 298 K obtained using a flow-type cryostat. The powder diffraction patterns were evaluated by the Rietveld method [57,58] using the FullProf package [59] and magnetic representational analysis was performed with the SARAh programme [60].

One crystal of RbCoCl$_3$ (7 x 14 x 8 mm$^3$; 2.5 g) was mounted in an aluminum can under helium gas and orientated in the (*hhl*) plane for single-crystal neutron scattering measurements. Elastic data were collected on the triple-axis spectrometer EIGER [61] at SINQ using a neutron wavelength of 2.360 Å with a double-focussing monochromator, a horizontal-focussing analyser and a 37 mm PG002 filter to suppress higher-order neutrons.

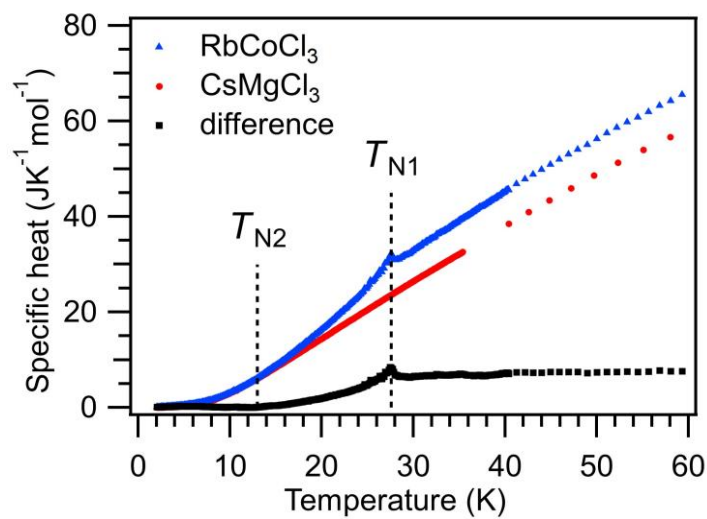

**Figure 2.** Specific heat of RbCoCl$_3$ and of the diamagnetic analogue CsMgCl$_3$ shown as a function of temperature. Their difference is shown in black.

## 3. Results and discussion

### *3.1. Magnetic specific heat*

The specific heat of RbCoCl$_3$, and of the diamagnetic analogue CsMgCl$_3$, is shown as a function of temperature in figure 2. Although the onset of 3D magnetic order is marked by a well-defined λ-anomaly at $T_{N1}$ = 27.6(2) K, it is clear from the difference curve in figure 2 that this ordering process releases only a very small fraction of the magnetic entropy. The recovery of this entropy up to temperatures in excess of 60 K indicates the presence of quasi-1D spin correlations well above $T_{N1}$, and we return to this topic in sections 3.2 and 3.4.2. In the specific heat there are no discernible features at the temperature, $T_{N2}$ = 13 K, where we show in section 3.4 how the magnetic structure changes from one ordered phase to another, suggesting that these changes have only minimal effects on the magnetic entropy.

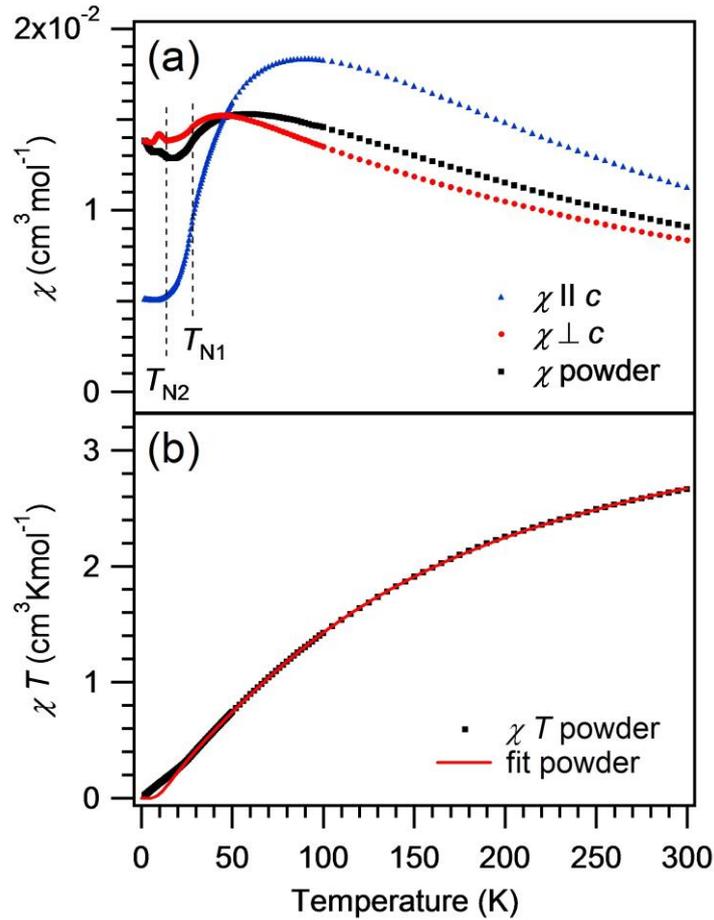

**Figure 3.** (a) Temperature-dependence of the magnetic susceptibility of RbCoCl$_3$ measured for a powder sample and for a single crystal in magnetic fields applied parallel and perpendicular to the hexagonal *c*-axis. The parallel susceptibility shows a broad maximum at 90 K that can be ascribed to the strong 1D AFM correlations. (b) Product $\chi T$ obtained from the powder data and fit according to equation (2).

### *3.2. Magnetic susceptibility*

For deeper insight into the anisotropic magnetic coupling between the Co$^{2+}$ ions, we performed temperature-dependent magnetic susceptibility measurements. The magnetisation of RbCoCl$_3$ measured at 1.85 K increases linearly in magnetic fields up to 5 T. Data for the susceptibility, $\chi(T)$, were collected in an applied magnetic field of 0.05 T for single-crystal and powder samples. Figure 3(a) displays features typical of a quasi-1D Ising spin system, reflecting both the magnetic anisotropy of the $Co^{2+}$ ions and the spatial anisotropy of the magnetic interactions in the hexagonal perovskite structure. The broad maxima appearing at quite different temperatures in the parallel and perpendicular magnetic susceptibilities again reflect predominantly 1D magnetic correlations associated with the chain direction. A crude estimate of the intrachain coupling constant, $J$, can be obtained from the temperature, $T_{max}$ = 90 K, of the maximum in the parallel susceptibility as $J \approx k_B T_{max}$ = 7.8 meV [62], where $k_B$ is the Boltzmann constant. Below we show how a more accurate estimate of $J$ may be extracted from the powder susceptibility and benchmark this with results known from INS.

The upper Néel temperature, $T_{N1}$, appears as an inflection point in the susceptibility curves of figure 3(a), indicating a qualitative change in the availability of magnetic fluctuations. Below $T_{N1}$ the susceptibility falls steeply, but below $T_{N2}$ the perpendicular and powder susceptibilities increase slightly. We comment that our data differ significantly from susceptibility results published in 1989 [63] and ascribe this to improved sample quality.

The quantity $\chi T$ for a polycrystalline sample decreases with decreasing temperature over the full range of our measurements, as shown in figure 3(b), and approaches zero at 1.9 K. The value at 300 K, 2.67 $cm^3$ K $mol^{-1}$, is significantly higher than the spin-only value of 1.87 $cm^3$ K $mol^{-1}$ (obtained for $g = 2$ and $S = 3/2$), indicating substantial orbital contributions. The small values at low temperatures are a consequence of both the AFM interactions and the zero-field splitting of the $Co^{2+}$ spins. The inverse susceptibility, $\chi^{-1}(T)$, shows linear behaviour only above 250 K, where a fit to the Curie-Weiss law yields the Curie constant $C$ = 4.09 $cm^3$ K $mol^{-1}$ and Curie temperature $\Theta$ = -160 K. The large and negative $\Theta$ value indicates again the presence of strong AFM interactions and can be used in a conventional susceptibility analysis to deduce the values of the magnetic interaction parameters. However, the $Co^{2+}$ ions in $RbCoCl_3$ adopt their high-spin state due to the weak crystal field of the $Cl^-$ ligands, and the combination of the resulting large zero-field splitting with the strong magnetic interactions complicates the modelling of $\chi(T)$ for $Co^{2+}$ compounds [64]. A comparison of different fitting methods [65] favours the empirical approach of Rueff *et al.* [66], according to which

$$\chi T = A e^{(\frac{-D}{k_B T})} + B e^{(\frac{-2J}{k_B T})}, \quad (1)$$

where $A + B = C$, the Curie constant, and $D$ is the zero-field splitting parameter. Subsequent modification to include an average $g$ value [67] led to the form

$$\chi T = N \left(\frac{\mu_B^2}{3 k_B}\right) g^2 \left[(C - \alpha) e^{(\frac{-D}{k_B T})} + \alpha e^{(\frac{-2J}{k_B T})}\right], \quad (2)$$

where $N$ is the number of ions per formula unit, $\mu_B$ the Bohr magneton and $\alpha$ a parameter fitting the ratio between the two contributions. Equation (2) has been applied successfully in a number of $Co^{2+}$ spin-chain materials [68], and for $RbCoCl_3$ we use the value of $C$ extracted from our data to deduce the parameters $g$ = 2.763(1), $D/k_B$ = 35.0(2) K and $2J/k_B$ = 162(1) K (with $\alpha$ = 2.74(1)). Both $g$ and $D$ have a well-defined physical meaning and this susceptibility fit provides accurate estimates of their values, which fall well within the ranges expected for $Co^{2+}$ chain compounds [64,67,69,70,71]. A physical interpretation for the single value of $J$ extracted from equation (2) is more complex, but for $RbCoCl_3$ we have measured the magnetic interaction parameters to high accuracy by INS [51]. The nearest-neighbour AFM Ising term, $2J_1/k_B$ = 136.6(2) K, is accompanied by a FM next-neighbour interaction, $2J_2/k_B$ = − 12.0(1) K, a Heisenberg interaction component, $2\varepsilon_1 J_1/k_B$ = 17.2(2) K and an interchain coupling $J_{nn}/k_B$ =

1.50(1) K. An effective value of 162 K is therefore fully consistent with the magnetic excitation spectrum, and hence our results help to benchmark the accuracy of $g$, $D$ and $J$ values extracted from the susceptibility analysis for systems where large single crystals are not available.

*3.3. Crystal structure*

The crystal structure of RbCoCl$_3$ was refined from neutron powder diffraction data collected on HRPT at temperatures between 1.5 K and 297 K. The diffraction pattern observed at $T = 35$ K, above both Néel temperatures, is presented in figure 4. It is straightforward to confirm that RbCoCl$_3$ is a hexagonal perovskite with space group P6$_3$/mmc, two formula units per unit cell and the atomic positions and separations shown in tables 2 and 3. These results coincide with the single crystal x-ray structure determined by Engberg and Soling [72] at room temperature (RT), where we obtain the lattice parameters $a = 7.0003(3)$ Å and $c = 5.9989(2)$ Å.

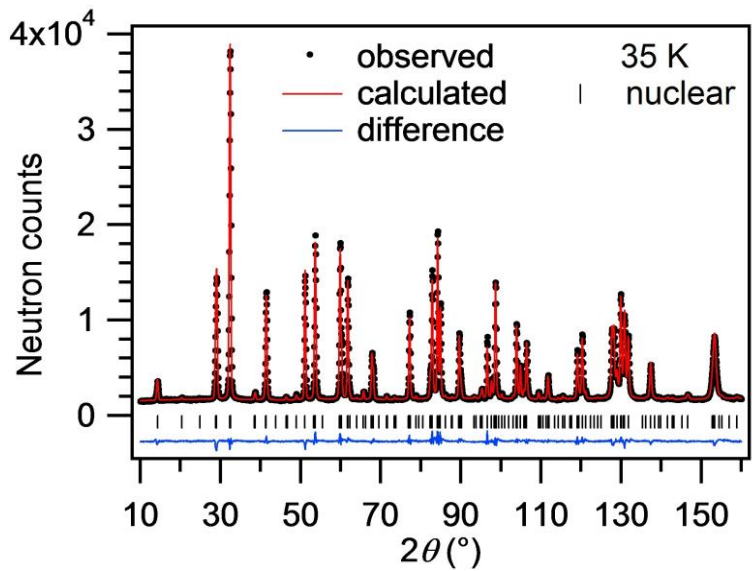

**Figure 4.** Neutron powder diffraction pattern of RbCoCl$_3$ at 35 K, collected on HRPT with $\lambda = 1.494$ Å.

**Table 2.** Atomic positions in RbCoCl$_3$ at 297 K and at 1.5 K. Space group P6$_3$/mmc, $Z = 2$.

| atom | site | x/a | y/b | z/c | T (K) |
|---|---|---|---|---|---|
| Rb | (2d) | 1/3 | 2/3 | 3/4 | |
| Co | (2a) | 0 | 0 | 0 | |
| Cl | (6h) | 0.16110(6) | 2x | 1/4 | 1.5 |
| Cl | (6h) | 0.16069(15) | 2x | 1/4 | 297 |
| Cl [72] | (6h) | 0.161(2) | 2x | 1/4 | RT |

**Table 3.** Interatomic separations in RbCoCl$_3$ at 1.5 K.

| atoms | separation (Å) | multiplicity |
|---|---|---|
| Rb – Cl | 3.4657(5) | 12 |
| Co – Cl | 2.4398(3) | 6 |
| Co – Co intrachain | 2.97547(3) | 2 |
| Co – Co interchain | 6.92992(7) | 6 |

The crystal structure is represented in figure 1: physically, $Rb^+$ and $Cl^-$ ions form hexagonal close-packed $RbCl_3$ layers in the *ab*-plane, which are stacked in an ABAB sequence along the *c*-axis, and the $Co^{2+}$ ions occupy the octahedral voids between planes. Chemically, $RbCoCl_3$ is composed of trigonally distorted $[CoCl_6]^{4-}$ octahedra, which share faces and form *c*-axis chains with a Co-Co separation of *c*/2. The chains are separated by $Rb^+$ ions, with Co-Co distances of *a* in the *ab*-plane, and these widely different separations are responsible for the predominantly 1D magnetic character.

The neutron powder diffraction data at all temperatures show no extra lines, and thereby confirm the phase purity of the sample. Upon cooling, there is no hint of a structural phase transition down to 1.5 K. The temperature-dependence of the measured lattice parameters, shown in the supplementary material (figure S1 and table S1), reveals that the lattice expands by 1.0 % and 0.8 %, respectively along the *a*- and *c*-axes, between 1.5 K and RT; the *c*-axis thermal expansion is expected to be slightly smaller due to the stronger ionic bonds in the $[CoCl_{6/2}]^-$ chains. The relative atomic positions are almost unaffected by the changing temperature, as shown by the *x*/*a* parameter of Cl in table 2. By inspection, the leading contribution to the intrachain magnetic interactions is from superexchange processes on all three Co-Cl-Co paths, while the shortest interaction pathway perpendicular to the chains is Co-Cl-Cl-Co, which includes a Cl-Cl van der Waals contact.

Sterically, these results indicate that the $Rb^+$ ionic radius, 1.72 Å [73] for coordination number 12, is large enough to maintain the hexagonal structure at least down to 1.5 K. This is also the case for $CsCoCl_3$, where the $Cs^+$ ionic radius of 1.88 Å is even closer to that of $Cl^-$ (1.81 Å). By contrast, the $Tl^+$ ion is only slightly smaller than $Rb^+$, with a radius of 1.70 Å, but cannot stabilise the hexagonal structure to low temperatures, and instead a series of structural phase transitions is observed in $TlCoCl_3$ below 165 K, in which the structure distorts towards lower symmetries and smaller volumes [42].

To exclude even the most minuscule low-temperature structural distortions in $RbCoCl_3$, we collected additional high-resolution synchrotron x-ray diffraction data on the X04SA-MS beamline at RT and 4 K. The high diffracted intensities, obtained typically with 2-3 minutes of counting time, yield estimated relative standard deviations varying from 5 x $10^{-3}$ for the background regions to 1 x $10^{-3}$ for the strongest peaks. The structural refinements performed using these data indeed confirm that there are no deviations from the 2H structure with space group $P6_3/mmc$, and nor do any additional Bragg peaks appear at low temperature. We draw particular attention to absence of structural peaks at the 2θ positions related to the magnetic propagation vectors, which are marked in figure S2. The origin of all the phase transitions observed by neutron diffraction may therefore be attributed to magnetic order.

### 3.4. Magnetic structures

By performing a series of neutron powder and single-crystal diffraction experiments at different temperatures, we have observed two magnetic phase transitions in $RbCoCl_3$, as figure 5(a) makes clear. At $T_{N1}$ = 28 K, the appearance of a number of strong Bragg peaks indicates that the system undergoes a transition to a phase of 3D AFM order, which we label the AFM1 phase. At $T_{N2}$ = 13 K, the appearance of additional weak Bragg peaks indicates a further modulation of the magnetic structure, and we label this second 3D AFM phase as AFM2. We present a detailed analysis of each phase in turn.

### 3.4.1. AFM1 structure below $T_{N1}$

To analyse the magnetic Bragg peaks appearing below 28 K, in figure 5(b) we show the difference between the neutron diffraction patterns obtained at 35 K and 18 K. Because the nuclear diffracted intensity is approximately constant in this temperature range, other than the small thermal lattice expansion, the difference pattern allows an accurate separation of the small magnetic contribution. The background scattering is lower in the AFM1 phase at 18 K than at 35 K, which is due to a partial redistribution of the paramagnetic diffuse scattering intensity (section 3.4.2) into the Bragg peaks.

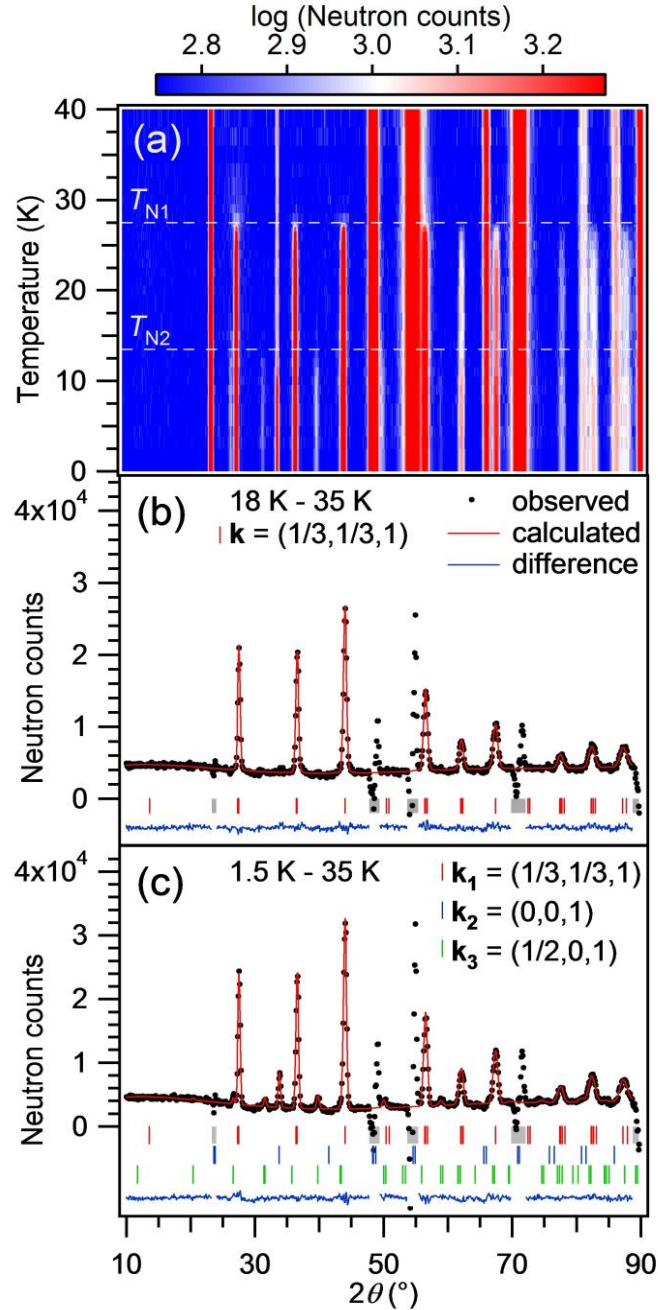

**Figure 5.** (a) Temperature-dependence of neutron powder diffraction patterns collected on DMC with $\lambda = 2.455$ Å. Additional magnetic Bragg peaks are observed below $T_{N1}$ and $T_{N2}$ due to 3D AFM order. Each diffraction pattern was refined individually and the fitted background was subtracted in order to increase the sensitivity to the weak magnetic peaks. (b) Difference pattern obtained by subtracting the 35 K data from the 18 K data. (c) Difference pattern obtained from 1.5 K − 35 K. A constant offset of 5000 counts was added in order to avoid negative count numbers. 2θ regions (grey) around strong nuclear reflections were excluded from the calculations because of thermal lattice expansion.

Prominent magnetic Bragg peaks are observed at 2θ values of 27.3°, 36.4°, 43.8°, 56.3° and 67.2°. All of these peaks can be indexed with the propagation vector **k** = (1/3,1/3,1), as shown in figure 5(b). RbCoCl$_3$ contains two Co$^{2+}$ ions in the unit cell, located at (0,0,0) and (0,0,0.5), and thus the $l = 1$ component of the **k**-vector indicates AFM order along the *c*-axis. The $h = k = 1/3$ components of the **k**-vector reveal AFM coupling in the *ab*-plane, leading to a $3a_0 \times 3a_0 \times c_0$ magnetic unit cell, which is represented in figure 6(a). Alternatively, the magnetic structure could be described using a $\sqrt{3}a_0 \times \sqrt{3}a_0 \times c_0$ cell, as indicated by the dashed line in figure 6(a), where the axes are rotated by 30° in the *ab*-plane. We have evaluated the possible spin structures based on a representational analysis [59], and the full results are presented in table S2 of the supplementary material, with the resulting magnetic structure refinements shown in table S3.

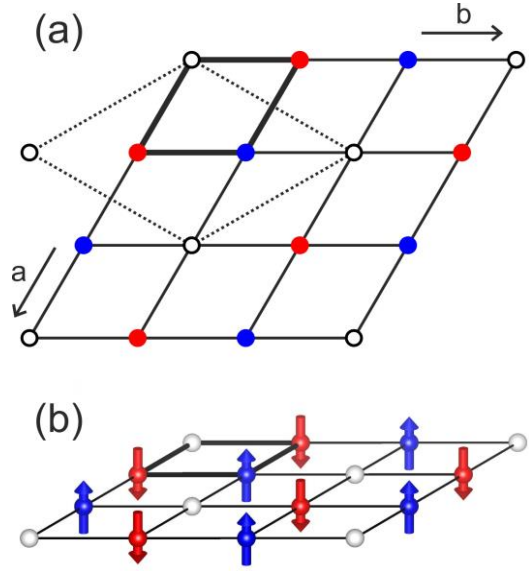

**Figure 6.** (a) Magnetic unit cell and spin structure of RbCoCl$_3$ in the AFM1 phase. Thick lines indicate the nuclear unit cell, thin lines the $3a_0 \times 3a_0 \times c_0$ magnetic unit cell and dashed lines the alternative description using a $\sqrt{3}a_0 \times \sqrt{3}a_0 \times c_0$ magnetic cell. Full circles represent Co$^{2+}$ ions at $z = 0$ with ordered magnetic moments $\mu = (0,0,\mu_z)$ (blue) and $(0,0,-\mu_z)$ (red), where $\mu_z = \sqrt{3}\mu_{z0}/2$. White circles represent ions that are "disordered" in the sense that their chains have no in-plane spin correlations. For Co$^{2+}$ ions at $z = 0.5$, the respective spins are reversed to $\mu = (0,0,-\mu_z)$ and $(0,0,\mu_z)$. (b) Schematic representation of the spin structure of RbCoCl$_3$ in the AFM1 phase that arises from **k** = (1/3,1/3,1), as specified in table 4.

Our analysis of the magnetic structure is performed by using FullProf [58]. It immediately excludes the irreducible representations (IRs) $\Gamma_3$ and $\Gamma_6$, both of which have $R_{mag} \approx 100\%$, meaning no reasonable fit is possible; at the qualitative level, both of these IRs contain a FM coupling along the *c*-axis, which contradicts the $l = 1$ component of the **k**-vector. The best solution, with $R_{mag} = 5.9\%$, is provided by $\Gamma_4$ and is a configuration in which the ordered component of the magnetic moment lies exclusively along the chain direction with amplitude $\mu_{z0} = 3.72(8)\mu_B$, as summarised in table 4. The moments have AFM coupling along the *c*-axis, in accordance with the phase shift of 180° ($\pi$) over the interatomic separation $c/2$. To supplement this part of the discussion we refer to the IR $\Gamma_5$, which has magnetic moments in the *ab*-plane with the modulation specified by **k** = (1/3,1/3,1). This type of spin structure cannot reproduce the observed magnetic diffraction intensities, yielding $R_{mag} = 60\%$, and in fact no combination of $\Gamma_4$ and $\Gamma_5$ can improve the fit obtained for $\Gamma_4$ alone. This result indicates the clear absence of ordered components of the magnetic moment in the *ab*-plane in the AFM1 phase and confirms their purely *c*-axis orientation.

**Table 4.** Result of the magnetic structure refinement based on the 18 K – 35 K difference data (DMC, $\lambda$ = 2.455 Å; $\chi^2$ = 1.869) for **k** = (1/3,1/3,1) and a $Co^{2+}$ ion located on site (2a) at (0,0,0). IR is the irreducible representation, $\mu_{x0}$, $\mu_{y0}$ and $\mu_{z0}$ are the amplitudes of the ordered magnetic components and $R_{mag}$ is the magnetic *R*-value.

| **k**-vector | IR | $\mu_{x0}$ ($\mu_B$) | $\mu_{y0}$ ($\mu_B$) | $\mu_{z0}$ ($\mu_B$) | $R_{mag}$ (%) |
|---|---|---|---|---|---|
| **k** | $\Gamma_4$ | 0 | 0 | 3.72(8) | 5.89 |

To discuss the interchain order in the AFM1 magnetic structure, we express the ordered magnetic moment on a $Co^{2+}$ ion at lattice site **r** = (*x*,*y*,*z*) as

$$\boldsymbol{\mu}(\boldsymbol{r}) = \boldsymbol{\mu}\cos(2\pi \boldsymbol{k}\cdot\boldsymbol{r} + \delta) \text{ with } \boldsymbol{k}\cdot\boldsymbol{r} = x/3 + y/3 + z, \quad (3)$$

where **μ** specifies the direction and absolute magnitude of the magnetic moment and $\delta$ is a phase offset that cannot be determined from powder diffraction data alone. In the AFM1 phase, an arbitrary value of $\delta$ means that the three different $Co^{2+}$ chains in the magnetic unit cell have different $\mu_z$ components. Here we discuss only the high-symmetry cases $\delta = 0$, which corresponds to magnetic space group P6$_3$'/m'cm', and $\delta = \pi/2$, which corresponds to P6$_3$/m'cm. However, any shift of $\delta$ by $\pi/3$ yields an equivalent magnetic structure with the origin shifted among the six $Co^{2+}$ positions of the magnetic unit cell. When $\delta = 0$, the ordered moments on the three different chains have magnitudes 1, -1/2 and -1/2 (in units of $\mu_{z0}$), while $\delta = \pi/2$ results in magnitudes of 0, -√3/2 and √3/2. The reduction of the ordered moment on certain sites that results from this modulation is usually described in terms of magnetic disorder. The corresponding diffuse elastic scattering, which we also observe in $RbCoCl_3$ (figures 7 and 8), is discussed in detail in section 3.4.2. This disorder interpretation has resulted in the AFM1 phase in $AMX_3$ compounds being referred to as the partially disordered AFM (PDAFM) phase.

While $\delta$ cannot be fixed only by powder diffraction, the magnetic order has direct repercussions for the nature of the spin excitations, and therefore we can supplement our analysis with the results for the spin excitation spectrum measured in the AFM1 phase by INS. In reference [51], the 18 K spectrum was interpreted as indicating a significant population (of order 40%) of Ising chains with no effective magnetic field, which appear as a scattering continuum, together with a population of chains experiencing net fields of twice and four times the interchain coupling constant (again in units of $\mu_{z0}$). These effective fields are deduced by summing over the six chains neighbouring any one chain, and performing this exercise for the $\delta = 0$ magnetic structure gives the result that one chain type has an effective field of $3J_{nn}$ while the other two have fields of $3J_{nn}/2$. By contrast, the $\delta = \pi/2$ magnetic structure yields 1/3 of the chains (those with ordered moment zero) whose effective field sums to zero and 2/3 of the chains in a field of $3\sqrt{3}J_{nn}/2$. Because only the $\delta = \pi/2$ case allows the appearance of a scattering continuum, as opposed to the Zeeman ladder of discrete bound states arising for an Ising chain in an effective magnetic field [51], we therefore interpret the INS spectrum as evidence that the interchain magnetic order in the AFM1 phase in $RbCoCl_3$ is the (0,-√3/2,√3/2) order set by $\delta = \pi/2$. This spin configuration, which is illustrated in figure 6, matches a number of analyses in the literature [43,44,45,46,74] and sets the magnetic space group P6$_3$/m'cm indicated in table 5.

In more detail, we stress that all chains in the system have strong intrachain Ising magnetic order. The designation "0" for 1/3 of the chains means that their spins are decorrelated from the spins of the other chains. The origin of this behaviour lies in the fact that the pure Ising model on a triangular lattice has an infinite degeneracy [75,76]. Our observation of **k** = (1/3,1/3,1) order already implies that the leading interaction acting to lift this degeneracy is a Heisenberg coupling of the *x* and *y* spin components, in contrast to the FM second-neighbour Ising interaction adopted in a number of studies [77,78,79,51]. Because these interactions are well below the resolution of any spectroscopic measurements, we take the low-temperature magnetic structures we have determined to provide the definitive statement of their consequences for the lifting of degeneracy. In the AFM1 phase, the solution is that 1/3 of the chains decorrelate to allow the remaining 2/3 to profit from interchain AFM order on an unfrustrated hexagonal, or honeycomb, lattice. Finally, the contribution to the excitation spectrum arising from chains experiencing an effective field is rather broad at 18 K (figures 3 and S5 of reference [51]), and would be fully consistent with our deduction that 2/3 of chains have an effective field of $3\sqrt{3}J_{nn}/2$.

Many AMX$_3$ compounds exhibit a **k** = (1/3,1/3,1) ordering vector in their AFM1 phase, as we showed in table 1. Depending on the single-ion anisotropy, the ordered moments can be oriented in the *xy*, *xz* or *z* directions. Only the Co$^{2+}$ compounds show ordered moments exclusively along the *z* direction, which is a consequence of their pronounced Ising anisotropy. To the extent that this can be gauged from literature data, all the AFM1 magnetic structures of Co$^{2+}$ compounds with an undistorted 2H structure are the same, despite the different descriptions used. The reported ordered moments range from 2.7 to 3.1μ$_B$ [43,44,45,46], and for RbCoCl$_3$ we have found a rather similar value of μ$_z$ = $\sqrt{3}$μ$_{z0}$/2 = 3.22(7)μ$_B$, while μ$_{z0}$ = 3.72(8)μ$_B$ approaches the spin-only value for high-spin Co$^{2+}$, $\sqrt{15}$μ$_B$ ≈ 3.87μ$_B$.

**Table 5.** Relative atomic positions, *x,y,z*, and ordered magnetic moments, **μ** (μ$_B$), of the Co$^{2+}$ ions on site (2a) in RbCoCl$_3$. For consistency the atomic positions and components of the ordered moments are indexed in the hexagonal basis for all three phases.

| Phase<br>Space group | **k**-vector<br>μ | Z | *x,y,z* | μ$_x$,μ$_y$,μ$_z$ | *x,y,z* | μ$_x$,μ$_y$,μ$_z$ |
|---|---|---|---|---|---|---|
| quasi-1D AFM<br>P6$_3$/mmc | | 2 | 0,0,0 | 0,0,0 | 0,0,0.5 | 0,0,0 |
| AFM1<br>P6$_3$/m'cm | **k** = (1/3,1/3,1)<br>μ$_{z0}$ = 3.72(8) | 6 | 0,0,0<br>1,0,0<br>1,1,0 | 0,0,0<br>0,0,-3.22<br>0,0,3.22 | 0,0,0.5<br>1,0,0.5<br>1,1,0.5 | 0,0,0<br>0,0,3.22<br>0,0,-3.22 |
| AFM2 | **k$_1$** = (1/3,1/3,1)<br>μ$_{z1}$ = 4.27(4)<br>**k$_2$** = (0,0,1)<br>μ$_{z2}$ = -1.08(10)<br>**k$_3$** = (1/2,0,1)<br>μ$_{x3}$ = 0.74(4),<br>μ$_{y3}$ = μ$_{x3}$/2 | 12 | 0,0,0<br>0,1,0<br>0,2,0<br>1,1,0<br>1,2,0<br>1,3,0 | 0.74,0.37,3.19<br>0.74,0.37,-3.19<br>0.74,0.37,-3.19<br>-0.74,-0.37,-3.19<br>-0.74,-0.37,3.19<br>-0.74,-0.37,-3.19 | 0,0,0.5<br>0,1,0.5<br>0,2,0.5<br>1,1,0.5<br>1,2,0.5<br>1,3,0.5 | -0.74,-0.37,-3.19<br>-0.74,-0.37,3.19<br>-0.74,-0.37,3.19<br>0.74,0.37,3.19<br>0.74,0.37,-3.19<br>0.74,0.37,3.19 |

### 3.4.2. AFM2 structure below $T_{N2}$ and diffuse scattering

Below the second ordering temperature, $T_{N2}$ = 13 K, we have seen in figure 5 that the magnetic structure becomes more complex. The observed Bragg peaks can be indexed to the simultaneous contributions of three different **k**-vectors. The magnetic peaks related to $\mathbf{k_1} = \mathbf{k} = (1/3,1/3,1)$ gain further intensity, which indicates an increase in the amplitude of the ordered moment from $\mu_{z0}$ = 3.72(8)$\mu_B$ at 18 K to $\mu_{z1}$ = 4.27(4)$\mu_B$ at 1.5 K, as shown in table 5. From the neutron powder diffraction data presented in figures 5(a) and 5(c), we have identified two additional propagation vectors, $\mathbf{k_2}$ = (0,0,1) and $\mathbf{k_3}$ = (1/2,0,1). The $\mathbf{k_2}$ contributions are most obvious from the nuclear-forbidden (1,1,1) peak at 2θ angle 33.8°, which shows a significant increase in intensity. The $\mathbf{k_3}$ contributions are weak but clearly visible, including at 31.6°, 39.8° and 58.8°.

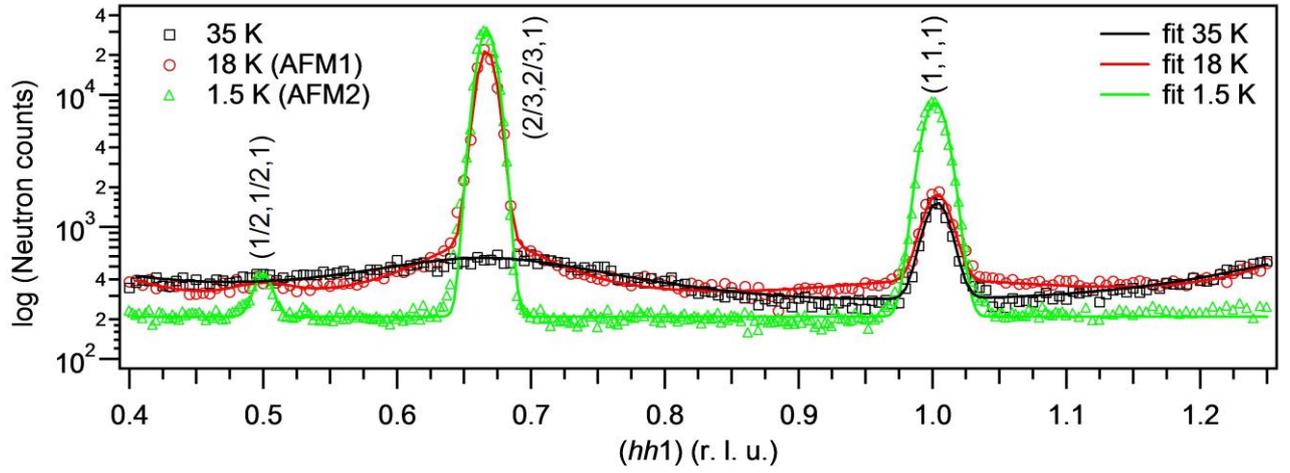

**Figure 7.** Neutron diffraction intensities measured in reciprocal lattice units (r.l.u.) along the ($hh$1) direction for a single crystal at three different temperatures using EIGER.

These **k**-vector assignments were verified in a neutron diffraction experiment using a single-crystal sample. Figure 7 shows data collected in elastic scans along the ($hh$1) direction for three characteristic temperatures, namely 35 K in the quasi-1D regime, 18 K in the AFM1 phase and 1.5 K in AFM2. The single-crystal data allow a separation of the elastic and diffuse contributions to the stronger magnetic peaks, the former being fitted by a Gaussian peak shape and the latter by Lorentzian functions. The background was assumed to be constant and temperature-independent. Figure 8 shows the complete temperature-dependence between 1.5 K and 60 K of the integrated intensities obtained for the (2/3,2/3,1), (1,1,1) and (1/2,1/2,1) magnetic reflections, and for their associated diffuse scattering.

A stepwise increase in the intensity of (2/3,2/3,1), a peak belonging to $\mathbf{k_1}$ = (1/3,1/3,1), is obvious from figure 8(a). The (1,1,1) reflection due to $\mathbf{k_2}$ = (0,0,1) gains intensity only below $T_{N2}$ = 13 K. The (1,1,1) nuclear reflection is forbidden in space group P6$_3$/mmc and the intensity observed above $T_{N2}$ in figure 7 is ascribed to a $\lambda$/2 contribution from the strong nuclear (2,2,2) peak, because the PG filter has a transmission of order 1% for high-energy neutrons. The rather weak peak at the (1/2,1/2,1) position shows the same temperature-dependence as the (1,1,1) peak. It belongs to $\mathbf{k_{3''}}$ = (-1/2,1/2,1), which is equivalent to $\mathbf{k_3}$ = (1/2,0,1), as discussed in more detail below. The onset of 3D magnetic order at $T_{N1}$ coincides with a maximum in the diffuse scattering. The diffuse component of $\mathbf{k_1}$ is observed well above $T_{N1}$ and reflects the presence of short-ranged 1D (i.e. intrachain) magnetic correlations. The diffuse component of $\mathbf{k_2}$ appears below 25 K, peaks at $T_{N2}$ and disappears at lower temperatures. Full saturation of the magnetic order is reached some way below $T_{N2}$, at approximately 8 K.

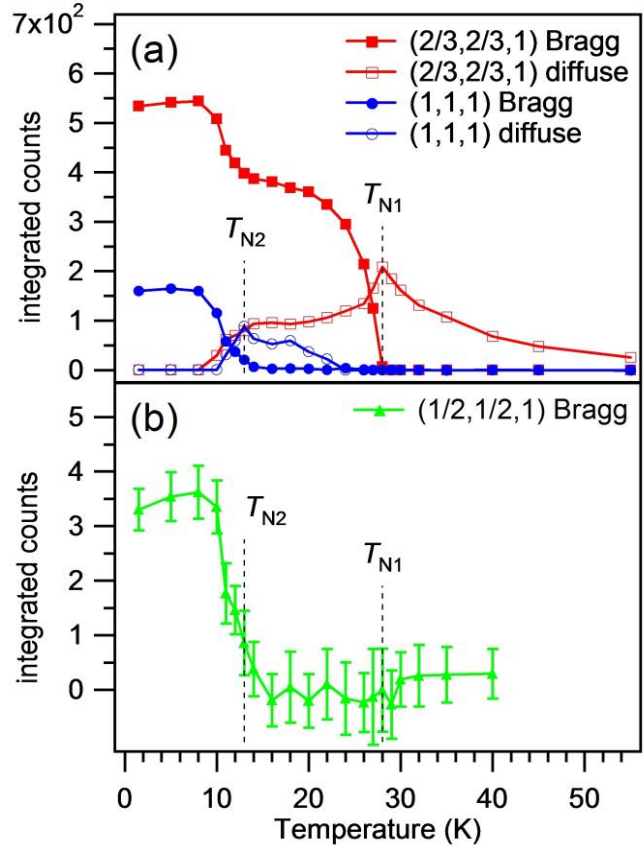

**Figure 8.** (a) Temperature-dependence of the integrated intensities of the (2/3,2/3,1) and (1,1,1) magnetic Bragg peaks, shown together with their diffuse scattering contributions. These peaks belong to $\mathbf{k_1}$ = (1/3,1/3,1) and $\mathbf{k_2}$ = (0,0,1), respectively. (b) Temperature-dependence of the (1/2,1/2,1) magnetic Bragg peak belonging to $\mathbf{k_3}$ (for reference see figure 10).

As discussed in section 3.4.1, in the regime between $T_{N1}$ and $T_{N2}$ the fluctuations arising from competing interactions in the triangularly frustrated *ab*-plane prevent full magnetic order. Because RbCoCl$_3$ maintains its hexagonal symmetry throughout the temperature range of our investigation, this frustration is not relieved by a structural distortion, as in the case of TlCoCl$_3$ [42]. Instead, this results below $T_{N2}$ in the simultaneous presence of three $\mathbf{k}$-vectors, which, as we will show, each add separate contributions to the total ordered magnetic moment. The additional $\mathbf{k}$-vectors enlarge the magnetic unit cell with respect to that of the AFM1 phase (figure 6), in that $\mathbf{k_3}$ doubles the magnetic unit cell below $T_{N2}$. The observed structure can be described by the orthorhombic (pseudo-C-centred) $\sqrt{3}a_0 \times 3a_0 \times c_0$ magnetic cell marked in figure 9(a) by the bold dashed line. The magnetic unit cell of the AFM2 phase contains 12 Co$^{2+}$ ions, whose ordered magnetic moments are calculated by analogy with equation (3) from the contributions of each of the three $\mathbf{k}$-vectors, as summarised in table 5.

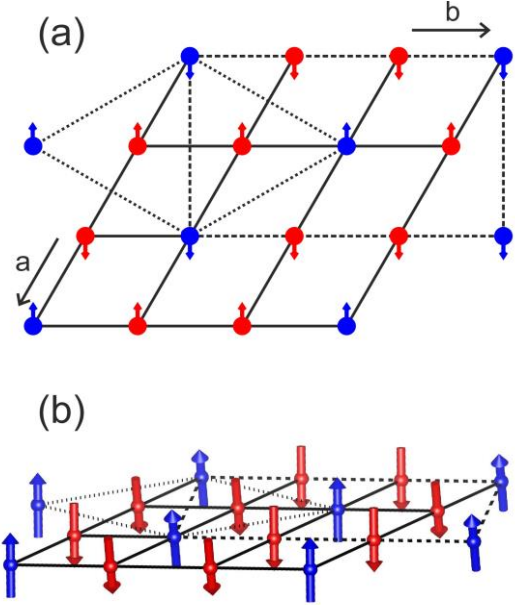

**Figure 9.** (a) Magnetic unit cell and spin structure of RbCoCl$_3$ in the AFM2 phase. The doubling of the hexagonal √3$a_0$ x √3$a_0$ x $c_0$ magnetic unit cell of the AFM1 phase (dotted line) results in an orthorhombic √3$a_0$ x 3$a_0$ x $c_0$ cell (dashed line). Circles represent axial $\mu_z$ (blue) and -$\mu_z$ (red) spin components of the Co$^{2+}$ ions at $z = 0$. Arrows indicate the in-plane spin components. All spins are reversed for Co$^{2+}$ ions at $z = 0.5$. (b) Schematic representation of the spin structure of RbCoCl$_3$ in the AFM2 phase that arises from $\mathbf{k_1} = (1/3,1/3,1)$, $\mathbf{k_2} = (0,0,1)$ and $\mathbf{k_3} = (1/2,0,1)$, as specified in table 5.

Because the transformation from a hexagonal to an orthorhombic cell results in multiple twinning, the weak $\mathbf{k_3} = (1/2,0,1)$ magnetic peaks are indexed using a set of three equivalent propagation vectors, a situation represented in figure 10 for the (*hk*1) plane in reciprocal space. This figure also shows the magnetic reflections referred to in figures 7 and 8, which are located along the diagonal of the nuclear Brillouin zone.

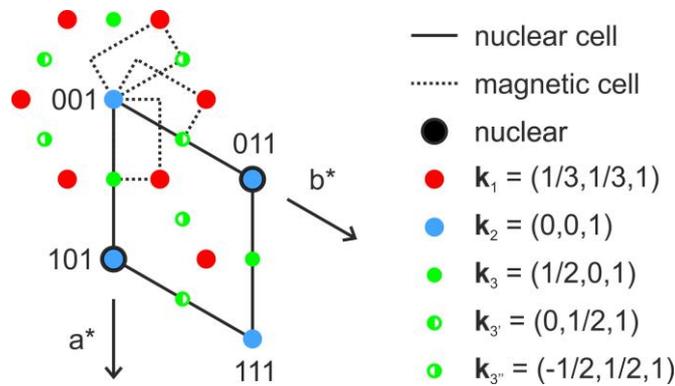

**Figure 10.** Section of the (*hk*1) plane, showing the nuclear and magnetic reciprocal cells of the AFM2 phase and the indexing of the observed propagation vectors.

We have performed a systematic representational analysis of the observed propagation vectors in the AFM2 phase, equivalent that for the AFM1 phase, and the results for both $\mathbf{k_2} = (0,0,1)$ and $\mathbf{k_3} = (1/2,0,1)$ are summarised in table S4 of the supplementary material. Because $\mathbf{k_2}$ contributes to the nuclear peak positions, which due to thermal expansion are only partially removed in the neutron powder diffraction difference pattern in figure 5(c), magnetic refinements were performed for both the 1.5 K and the 1.5 K – 35 K data sets. Following the discussion for $\mathbf{k_1} = (1/3,1/3,1)$ in the AFM1 phase, the IRs $\Gamma_3$ and $\Gamma_9$ of $\mathbf{k_2}$ show a FM coupling along the $c$-axis (table S2 of the supplementary material). This contradicts the $l = 1$ component of $\mathbf{k_2}$ and the observed magnetic intensities cannot be reproduced. $\Gamma_7$ and $\Gamma_{11}$ both show an AFM coupling along the $c$-axis, with magnetic moments respectively along and perpendicular to $c$; with $R_{mag}$= 8.45%, $\Gamma_7$ clearly yields better agreement and refines consistently to a contribution of $\mu_{z2}$ = -1.08(10)$\mu_B$ for both data sets.

Turning to $\mathbf{k_3}$, again the $l = 1$ component excludes the spin configurations $\Gamma_3\psi_3$, $\Gamma_5$ and $\Gamma_7\psi_5$, which have a FM coupling along the $c$-axis (table S2 of the supplementary material) and fail to reproduce the observed magnetic intensities. Among the candidates with AFM coupling along $c$, $\Gamma_1$ ($R_{mag}$= 59.6%) and $\Gamma_3\psi_2$ ($R_{mag}$= 47.5%) have ordered moments lying in the $ab$-plane while for $\Gamma_7\psi_6$ ($R_{mag}$= 23.0%) they are aligned along $c$. Due to the relatively weak contributions of the $\mathbf{k_3}$-related magnetic ordering to the overall magnetic intensity, all of these $R_{mag}$ values are higher than for $\mathbf{k_1}$ and $\mathbf{k_2}$. Nevertheless, we comment that the 1.5 K data set measured on DMC at long wavelengths and the two data sets obtained from HRPT using two different wavelengths all led to consistent results. To decide among these possibilities, as in section 3.4.1 we consider the excitation spectrum measured in the AFM2 phase [51].

In the absence of $\mathbf{k_3}$, $\mathbf{k_1}$ and $\mathbf{k_2}$ together produce an ordered state with all moments of equal magnitude. In contrast to the AFM1 phase, where $\delta = \pi/2$, in the AFM2 phase we find that $\delta = 0$, yielding spin magnitudes of 1, -1/2, -1/2, in units of $\mu_{z1}$= 4.27(4)$\mu_B$, aligned along the $c$-axis. The sum with the constant $\mu_{z2}$ component, which is almost identically equal to $\mu_{z1}/4$, yields spin magnitudes of 3/4, -3/4, -3/4 in units of $\mu_{z1}$. This ordered state therefore has equal moments, $\mu_z = 3.19(14)\mu_B$, aligned in an up-down-down configuration in each $ab$-plane, as shown in figure 9; because the majority spins lie on a hexagonal lattice, this state was referred to in Ref. [51] as a "honeycomb" configuration, and its moment imbalance is of course offset by the neighbouring planes due to the strong AFM chain interactions. We comment that the ordered moment $\mu_z = 3.19(14)\mu_B$ determined in the AFM2 phase is within the error bars equal to the value $\mu_z = 3.22(7)\mu_B$ obtained for AFM1, but stress that in the AFM2 phase all chains are correlated within the hexagonal plane.

The IR $\Gamma_7\psi_6$ of $\mathbf{k_3}$ would contribute an additional out-of-plane ordered component, $\mu_{z3} = 0.92(4)\mu_B$, which would establish a state with 1/2 of the sites having an ordered moment $\mu_{z,eff} = 4.11(18)\mu_B$ and 1/2 having 2.27(18)$\mu_B$; with both types orientated 1/3 up and 2/3 down in a single plane, 2/3 of the chains would experience a weak effective field of approximately $0.6J_{nn}$ while 1/3 experience large effective fields up to $6.5J_{nn}$ (both in units of $\mu_z = 3.19\mu_B$). The candidates with only in-plane moment contributions have the strong advantage of producing precisely the "honeycomb" configuration of the out-of-plane moments considered in reference [51] to interpret the spectroscopic data. For the in-plane components, both $\Gamma_3\psi_2$ and $\Gamma_1$ yield ordered moments of approximately 0.6$\mu_B$, aligned respectively along the orthorhombic $a$- or $b$-axes, which would be indistinguishable in the hexagonal parent cell [80]. Here we focus on $\Gamma_3\psi_2$, because its $R_{mag}$ and $\chi^2$ values are somewhat smaller than those of $\Gamma_1$ (table S4): it contributes the in-plane structure represented in figure 9, in which alternating lines of ordered magnetic components, $\mu_x = \sqrt{3}\mu_{x3}/2 = 0.64(3)\mu_B$, are orientated along the orthorhombic $a$-axis and have a staggered $c$-axis alignment. This causes every Ising chain to have its moments tilted 11.3° off the $c$-axis, but to remain mutually AFM, and to experience a staggered transverse effective field of approximately $0.4J_{nn}$ (again in units of $\mu_z = 3.19\mu_B$).

**Table 6.** Results of the magnetic structure refinement based on the 1.5 K – 35 K difference data (DMC, $\lambda$ = 2.455 Å; $\chi^2$ = 2.91) for $\mathbf{k_1}$ = (1/3,1/3,1), $\mathbf{k_2}$ = (0,0,1) and $\mathbf{k_3}$ = (1/2,0,1) for a $Co^{2+}$ ion located on site (2a) at (0,0,0). $\mu_{xi}$, $\mu_{yi}$ and $\mu_{zi}$ are the amplitudes of the ordered magnetic components for i = 1,2,3.

| $\mathbf{k}$-vector | IR | $\mu_{xi}$ ($\mu_B$) | $\mu_{yi}$ ($\mu_B$) | $\mu_{zi}$ ($\mu_B$) | $R_{mag}$ (%) |
|---|---|---|---|---|---|
| $\mathbf{k_1}$ | $\Gamma_4$ | 0 | 0 | 4.27(4) | 3.82 |
| $\mathbf{k_2}$ | $\Gamma_7$ | 0 | 0 | -1.08(10) | 8.45 |
| $\mathbf{k_3}$ | $\Gamma_3\psi_2$ | 0.74(4) | $\mu_{x3}/2$ | 0 | 47.5 |

To decide between these two scenarios, the spectroscopic data at 4 K (figure 2 of reference [51]) clearly indicate strong continuum scattering, from approximately 2/3 of the chains, while the remaining 1/3 provide the scattering fingerprint of a strongly confined Zeeman ladder ($6J_{nn}$). In $\Gamma_7\psi_6$, the field on 2/3 of the chains is weak but finite, and explicitly longitudinal, meaning that it would act to create Zeeman-ladder bound states. Although this field is weak, the alteration of the spectrum from a continuum to discrete is a qualitative effect, and it is difficult to argue with certainty that any field is sufficiently weak as to be ineffective in creating a spectrum with some Zeeman-ladder features. By contrast, for $\Gamma_3\psi_2$ the traverse-field Ising model is a problem that remains exactly soluble in terms of free fermions, meaning that the scattering continuum of the chains in zero effective longitudinal field would be preserved in a small transverse field. Given the uncertainties in parameters fitted in reference [51], the small quantitative changes to the dispersion caused by the transverse field would not be discernible. On the basis of this difference between qualitative or quantitative modification of the spectrum by the low-temperature magnetic structure, we believe that $\Gamma_3\psi_2$ presents the scenario most consistent with the experimental evidence, despite the relative $R_{mag}$ values. The results for this refinement of the AFM2 magnetic structure are illustrated in figure 9 and summarised in table 6, with the associated model reproducing the intensities reported in figures 5(a), 5(c) and 7 to high accuracy.

We comment here that the individual refinement of the three contributions, due to $\mathbf{k_1}$, $\mathbf{k_2}$ and $\mathbf{k_3}$, was valuable in allowing us to unravel the components of this complex magnetic structure. It is feasible in RbCoCl$_3$ because each $\mathbf{k}$-vector contributes to a unique set of magnetic ($h,k,l$) values, with no overlap between the three groups. The results we obtained in this way were verified against a single-phase refinement with 12 Co atoms in the $\sqrt{3}a_0$ x $3a_0$ x $c_0$ magnetic cell.

To summarise, our systematic analysis of the AFM2 magnetic structure of RbCoCl$_3$ reveals for the first time the simultaneous presence of three $\mathbf{k}$-vectors. The interplay of these ordering vectors within and perpendicular to the *ab*-plane results in a magnetic structure with all moments of equal magnitude, and orientated 11.3° away from the *c*-axis. Although the overall features of the magnetic structure of RbCoCl$_3$ are similar to those reported for CsCoBr$_3$ [46] and CsCoCl$_3$ [43,44], it differs from each in certain respects. A comparison of our diffraction data in figure 5 to the data for CsCoBr$_3$ presented by Yelon *et al.* (1975) [46] shows that the magnetic unit cell of CsCoBr$_3$ is the same orthorhombic cell described here for RbCoCl$_3$ (figure 9). Both have the same out-of-plane order, and differ only in that the in-plane components have a relative rotation of 90°. This out-of-plane order, meaning the pattern governed by $\mathbf{k_1}$ and $\mathbf{k_2}$ that leads to equal-sized $\mu_z$ components orientated up-down-down in the hexagonal unit cell, also matches the "ferrimagnetic" low-temperature structure found in CsCoCl$_3$ [43,44]. (We recall that the term "ferrimagnetic" is misleading, because the dominant interaction is the 1D AFM coupling along the chains and this causes the net magnetisation to vanish, hence our preference for the term "honeycomb" to describe the geometry of the 2:1 ordered magnetic structure in each plane.)

## 4. Conclusions

RbCoCl$_3$ is an AFM Ising spin-chain material, based on face-sharing [CoCl$_{6/2}$]$^-$ octahedra, whose short-ranged 1D magnetic correlations peak around 80 K and dominate the measured magnetic specific heat and parallel susceptibility. The atomic structure of RbCoCl$_3$ maintains its hexagonal symmetry down to the lowest temperatures, meaning that the in-plane magnetic interactions are geometrically frustrated, and thus give rise to complex forms of low-temperature magnetic order. Below $T_{N1}$ = 28 K, a phase of "partial" 3D order, AFM1, is established, and below $T_{N2}$ = 13 K we find a pattern of complex but complete 3D order with three propagation vectors, the AFM2 phase; both Néel temperatures appear only as small features in the magnetic specific heat and susceptibility. AFM1 is described by the propagation vector $\mathbf{k} = \mathbf{k_1} = (1/3,1/3,1)$, which is manifest as a $(0,-\sqrt{3}/2,\sqrt{3}/2)$ ordering of the Ising chains in which every third chain is uncorrelated with its neighbours, allowing these to form an unfrustrated ordered phase on the hexagonal lattice. AFM2 is characterised by the additional propagation vectors $\mathbf{k_2}$ = (0,0,1) and $\mathbf{k_3}$ = (1/2,0,1), where $\mathbf{k_2}$ adds to the out-of-plane ordered moment to form a "honeycomb" magnetic configuration of equal moment amplitudes in a 2:1 ratio in each plane, while $\mathbf{k_3}$ contributes a weak and staggered in-plane moment. We have identified both the AFM1 and AFM2 structures among a number of competing candidates by ensuring their consistency with the basic features of the magnetic excitation spectra measured at different temperatures by neutron spectroscopy. Our results serve in turn as a starting point for more detailed studies of the confined and deconfined domain-wall excitations of the Ising chain in RbCoCl$_3$, where the material parameters allow the application of real or effective, longitudinal or transverse and uniform or staggered fields, as well as the possibility of thermal and other forms of disorder.

## Acknowledgments

We thank D. Biner for technical support, S. Decurtins for providing the SQUID magnetometer and M. Enderle, V. Pomjakushin and A. Wills for helpful discussions. This work is based on experiments performed at the Swiss spallation neutron source, SINQ, at the Paul Scherrer Institute (PSI, Villigen, Switzerland) and at the X04SA-MS beamline of the SLS synchrotron at PSI. We are grateful to the Swiss National Science Foundation (SNF) for financial support under the projects 200020-150257 and 200020-172659.

**Supplementary Material to accompany the manuscript**

# Magnetic order in the quasi-one-dimensional Ising system RbCoCl$_3$

N P Hänni, D Sheptyakov, M Mena, E Hirtenlechner, L Keller, U Stuhr, L-P Regnault, M Medarde, A Cervellino, Ch Rüegg, B Normand and K W Krämer

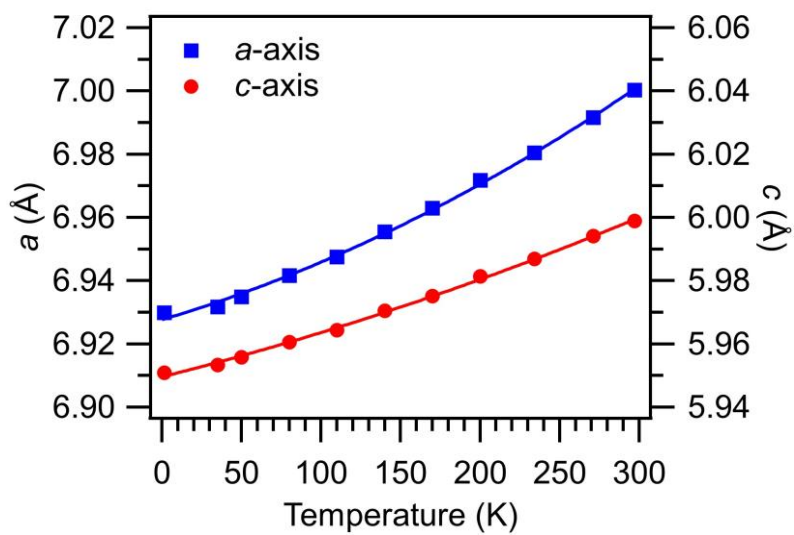

**Figure S1.** Temperature-dependence of the lattice parameters of RbCoCl$_3$ between 1.5 K and room temperature (RT). The lines are guides to the eye (3$^{rd}$ order polynomial).

**Table S1.** Temperature-dependence of the lattice parameters of RbCoCl$_3$.

| T (K) | a (Å) | c (Å) |
|---|---|---|
| 297 | 7.0003(3) | 5.9989(2) |
| RT[a] | 6.999(1) | 5.996(1) |
| 271 | 6.9916(2) | 5.9941(2) |
| 234 | 6.9804(2) | 5.9869(2) |
| 200 | 6.9718(2) | 5.9814(2) |
| 170 | 6.9629(2) | 5.9751(2) |
| 140 | 6.9555(2) | 5.9704(2) |
| 110 | 6.9475(2) | 5.9644(2) |
| 80 | 6.9416(2) | 5.9606(2) |
| 50 | 6.9349(2) | 5.9558(2) |
| 35 | 6.9317(1) | 5.9534(1) |
| 1.5 | 6.9299(1) | 5.9510(1) |

a) Engberg Å and Soling H 1967 *Acta Chem. Scand.* **21** 168

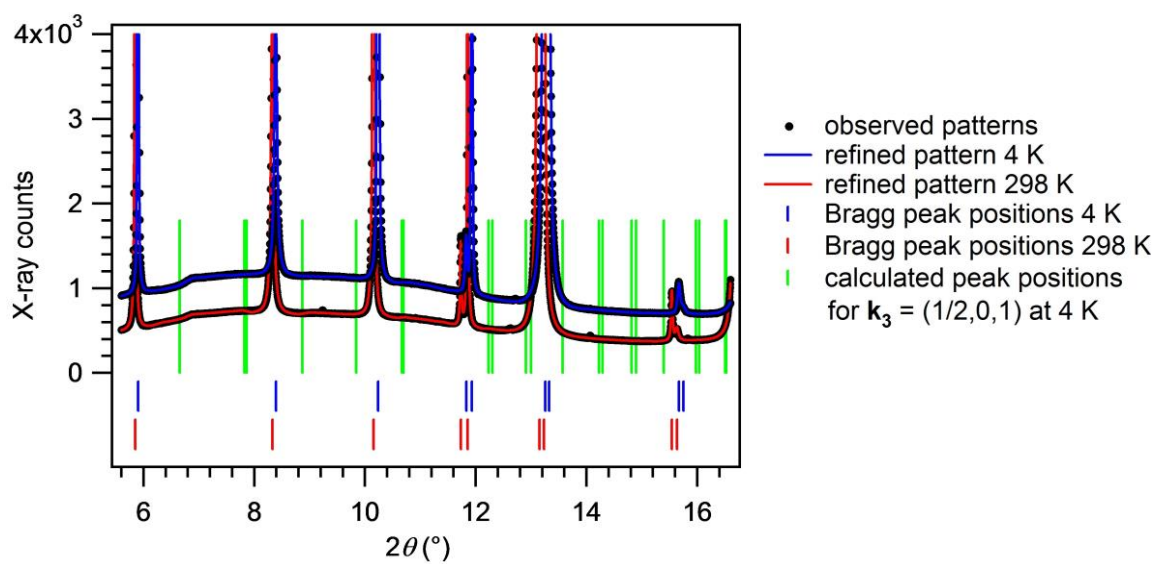

**Figure S2.** X-ray powder diffraction patterns of RbCoCl$_3$ at 298 K and 4 K, collected on the MS-powder beamline with λ = 0.62041 Å. Observed data, Rietveld refinements and calculated peak positions for both temperatures are shown along with the positions (vertical lines) for the hypothetical diffraction peaks that would correspond to the **k$_3$** = (1/2,0,1) modulation of the structure. The hypothetical peaks are absent even at the lowest temperature.

**Table S2.** Basis vectors (BVs) for space group $P6_3/mmc$ with the propagation vectors $\mathbf{k_1} = (1/3,1/3,1)$, $\mathbf{k_2} = (0,0,1)$ and $\mathbf{k_3} = (1/2,0,1)$. The decompositions of the magnetic representations into irreducible representations (IRs) are given for the $Co^{2+}$ site (2a) with atomic positions (0,0,0) (atom 1) and (0,0,1/2) (atom 2). The calculations were performed with the programme SARAh.[a]

| k-vector | IR | BV | atom | $m_a$ | $m_b$ | $m_c$ |
|---|---|---|---|---|---|---|
| (1/3,1/3,1) | $\Gamma_3$ | $\psi_1$ | 1 | 0 | 0 | 1 |
| | | | 2 | 0 | 0 | 1 |
| | $\Gamma_4$ | $\psi_2$ | 1 | 0 | 0 | 1 |
| | | | 2 | 0 | 0 | -1 |
| | $\Gamma_5$ | $\psi_3$ | 1 | 1 | 0 | 0 |
| | | | 2 | -1 | 0 | 0 |
| | | $\psi_4$ | 1 | 1 | 2 | 0 |
| | | | 2 | -1 | -2 | 0 |
| | $\Gamma_6$ | $\psi_5$ | 1 | 1 | 0 | 0 |
| | | | 2 | 1 | 0 | 0 |
| | | $\psi_6$ | 1 | 1 | 2 | 0 |
| | | | 2 | 1 | 2 | 0 |
| (0,0,1) | $\Gamma_3$ | $\psi_1$ | 1 | 0 | 0 | 1 |
| | | | 2 | 0 | 0 | 1 |
| | $\Gamma_7$ | $\psi_2$ | 1 | 0 | 0 | 1 |
| | | | 2 | 0 | 0 | -1 |
| | $\Gamma_9$ | $\psi_3$ | 1 | 1 | 0 | 0 |
| | | | 2 | 1 | 0 | 0 |
| | | $\psi_4$ | 1 | 1 | 2 | 0 |
| | | | 2 | 1 | 2 | 0 |
| | $\Gamma_{11}$ | $\psi_5$ | 1 | 1 | 0 | 0 |
| | | | 2 | -1 | 0 | 0 |
| | | $\psi_6$ | 1 | 1 | 2 | 0 |
| | | | 2 | -1 | -2 | 0 |
| (1/2,0,1) | $\Gamma_1$ | $\psi_1$ | 1 | 0 | 1 | 0 |
| | | | 2 | 0 | -1 | 0 |
| | $\Gamma_3$ | $\psi_2$ | 1 | 2 | 1 | 0 |
| | | | 2 | -2 | -1 | 0 |
| | | $\psi_3$ | 1 | 0 | 0 | 1 |
| | | | 2 | 0 | 0 | 1 |
| | $\Gamma_5$ | $\psi_4$ | 1 | 0 | 1 | 0 |
| | | | 2 | 0 | 1 | 0 |
| | $\Gamma_7$ | $\psi_5$ | 1 | 2 | 1 | 0 |
| | | | 2 | 2 | 1 | 0 |
| | | $\psi_6$ | 1 | 0 | 0 | 1 |
| | | | 2 | 0 | 0 | -1 |

a) Wills A S 2000 *Physica B* **276** 680

**Table S3**. Refinements for all possible IRs (table S2) for $\mathbf{k}_1 = (1/3,1/3,1)$ in the AFM1 phase based on data from the temperature difference 18 K – 35 K (DMC, $\lambda = 2.455$Å). The basis vectors $\psi_3$ and $\psi_4$ for the IR $\Gamma_5$, and $\psi_5$ and $\psi_6$ for $\Gamma_6$, produce identical diffraction patterns and differ only in moment orientation, as a result of which moment amplitudes are shown for only the first one of each pair.

| IR | chain order | $\mu_{x0}$ ($\mu_B$) | $\mu_{y0}$ ($\mu_B$) | $\mu_{z0}$ ($\mu_B$) | $R_{mag}$ (%) | $\chi^2$ |
|---|---|---|---|---|---|---|
| $\Gamma_3$ | FM | 0 | 0 | 0.6(4) | 100 | 176 |
| **$\Gamma_4$** | **AFM** | **0** | **0** | **3.72(8)** | **5.89** | **1.87** |
| $\Gamma_5\psi_3 \equiv \Gamma_5\psi_4$ | AFM | 0 | -2.48(16) | 0 | 60.1 | 64.2 |
| $\Gamma_6\psi_5 \equiv \Gamma_6\psi_6$ | FM | 0 | -1.0(5) | 0 | 99.7 | 174 |

**Table S4**. Refinements for all possible IRs (table S2) for $\mathbf{k}_1 = (1/3,1/3,1)$, $\mathbf{k}_2 = (0,0,1)$ and $\mathbf{k}_3 = (1/2,0,1)$ in the AFM2 phase based on data from the temperature difference 1.5 K – 35 K (DMC, $\lambda = 2.455$Å). Again only the first of each pair is shown when refinements are equivalent.

$\mathbf{k}_1 = (1/3,1/3,1)$

| IR | chain order | $\mu_{x1}$ ($\mu_B$) | $\mu_{y1}$ ($\mu_B$) | $\mu_{z1}$ ($\mu_B$) | $R_{mag}$ (%) | $\chi^2$ |
|---|---|---|---|---|---|---|
| $\Gamma_3$ | FM | 0 | 0 | 0.8(4) | 101 | 237 |
| **$\Gamma_4$** | **AFM** | **0** | **0** | **4.27(4)** | **3.82** | **2.35** |
| $\Gamma_5\psi_3 \equiv \Gamma_5\psi_4$ | AFM | 0 | -2.94(18) | 0 | 56.9 | 81.3 |
| $\Gamma_6\psi_5 \equiv \Gamma_6\psi_6$ | FM | 0 | -1.2(5) | 0 | 100 | 234 |

$\mathbf{k}_2 = (0,0,1)$

| IR | chain order | $\mu_{x2}$ ($\mu_B$) | $\mu_{y2}$ ($\mu_B$) | $\mu_{z2}$ ($\mu_B$) | $R_{mag}$ (%) | $\chi^2$ |
|---|---|---|---|---|---|---|
| $\Gamma_3$ | FM | | | | no convergence | |
| **$\Gamma_7$** | **AFM** | **0** | **0** | **-1.08(10)** | **8.45** | **2.354** |
| $\Gamma_9\psi_3 \equiv \Gamma_9\psi_4$ | FM | | | | no convergence | |
| $\Gamma_{11}\psi_5 \equiv \Gamma_{11}\psi_6$ | AFM | 0 | -0.88(8) | 0 | 8.60 | 2.328 |

$\mathbf{k}_3 = (1/2,0,1)$

| IR | chain order | $\mu_{x3}$ ($\mu_B$) | $\mu_{y3}$ ($\mu_B$) | $\mu_{z3}$ ($\mu_B$) | $R_{mag}$ (%) | $\chi^2$ |
|---|---|---|---|---|---|---|
| $\Gamma_1$ | AFM | 0 | -0.60(3) | 0 | 58.6 | 3.13 |
| **$\Gamma_3\psi_2$** | **AFM** | **0.74(4)** | **0.37(2)** | **0** | **47.5** | **2.91** |
| $\Gamma_3\psi_3$ | FM | 0 | 0 | 0.14(7) | 98.5 | 4.37 |
| $\Gamma_5$ | FM | 0 | -0.12(9) | 0 | 101 | 4.37 |
| $\Gamma_7\psi_5$ | FM | 0.48(8) | 0.24(4) | | 68.9 | 4.18 |
| $\Gamma_7\psi_6$ | AFM | 0 | 0 | 0.92(4) | 23.0 | 2.35 |